\documentclass[aps,prstab,preprint,groupedaddress]{revtex4}
\usepackage{graphics,epsfig}

\begin{document}

\title{Many-Body Theory for Multi-Agent Complex Systems}


\author{Neil F. Johnson$^*$, David M.D. Smith$^+$ and Pak Ming Hui$^{**}$}
\affiliation{$^*$Physics Department, Oxford University, Oxford, OX1 3PU, UK\\
$^+$Mathematics Department, Oxford University, Oxford, OX1 2EL, UK\\
$^{**}$Department of Physics, Chinese University of Hong Kong, Hong Kong}

\date{\today}

\begin{abstract} Multi-agent complex systems comprising populations of
decision-making particles, have wide application across the biological,
informational and social sciences. We uncover a formal analogy between these
systems' time-averaged dynamics and conventional
many-body theory in Physics. Their behavior is dominated by the
formation of `Crowd-Anticrowd' quasiparticles.  For the specific example of the
Minority Game, our formalism yields analytic expressions which are in excellent
agreement with numerical simulations. \vskip0.5in
\end{abstract}

\maketitle

Multi-agent simulations are currently being used to study the
dynamical behavior within a wide variety of Complex Systems
\cite{casti}. Within these simulations, $N$ decision-making
particles or {\em agents} (e.g. commuters, traders, computer
programs, cancer/normal cells, guerillas \cite{econo,rob,book})
repeatedly compete with each other for some limited global
resource (e.g. road space, best buy/sell price, processing time,
nutrients and physical space, political power) using sets of rules
which may differ between agents and may change in time. The
population is therefore competitive, heterogeneous and adaptive. A
simple minimal model which has generated more than one hundred
papers since 1998, is the Minority Game (MG) of Challet and Zhang
\cite{econo,book,others,savit,emg,us,paul,chau}. Numerical
simulations can yield fascinating results -- however it is very
hard to develop a general yet analytic theory of such multi-agent
systems.

Given that a multi-agent population is a many-body interacting
system with the {\em additional} complication of the particles
being decision-making, one wonders whether it might be possible to
develop a generalized `many-body' theory of such systems. This is
a daunting task since the success of conventional many-body theory
relies on the fundamental physical particles having a relatively
simple internal configuration space (e.g. spin) which is identical
for each particle -- moreover, the particle-particle interactions
are time-independent \cite{mahan}. By contrast, each
decision-making particle lives in a complex configuration space
represented by the information it receives and the particular
strategies which it happens to possess. In addition, the
agent-agent interactions generally evolve in time and depend on
prior history. However, it is {\em precisely} these difficulties
which make this problem  so interesting to a theoretical
physicist. In addition to the important real-world applications
listed above, such a generalized many-body theory could be applied
to physical systems where internal degrees of freedom can be
created artificially. An interesting technological example
concerns an array of interacting or interconnected nanostructures,
where each nanostructure has its own active defects which respond
to the collective actions of the others \cite{defects}.

Here we propose a general many-body-like formalism for these
complex $N$-body systems. Inspired by conventional many-body
theory \cite{mahan}, it is based around the accurate description
of the correlations between groups of agents. We show that the
system's fluctuations will in general be dominated by the
formation of crowds, and in particular the anti-correlation
between a given crowd and its mirror-image (i.e. `anticrowd'). The
formalism, when applied to the MG, yields a set of analytic
results which are in excellent agreement with the numerical
findings of Savit {\em et al.} (see Fig. 1) \cite{savit}. We note
that there have been many other MG theories proposed to date
\cite{others}, yet none of these has provided an analytic
description of the Savit-curve \cite{savit} over the entire
parameter space. In what follows, we do {\em not} restrict
ourselves to the MG -- for example, our formalism can be easily
generalized to {\em multiple} options \cite{chau}. We hope that
our results stimulate many-body physicists to investigate
transferring their techniques to these more general systems and
real-world applications.

Consider $N$ agents (e.g. commuters) who repeatedly decide between
two actions at each timestep $t$ (e.g. $+1/-1 \equiv$ take route
A/B) using their individual $S$ strategies.  Our formalism will
apply to  a wide variety of multi-agent games since it is
reasonably insensitive to the game's rules concerning
strategy-choice, rewards, and the definition of the winning group.
The agents have access to a common information source $\mu(t)$
which they use to decide actions. This information may be global
or local, correct or wrong, internally or externally generated.
Each strategy, labelled $R$, comprises a particular action for
each $\mu(t)\in \{\mu(t)\}$, and the set of possible strategies
constitutes a strategy space $\Theta\equiv \{R\}$.  The strategy
allocation among agents can be described in terms of a rank-$S$
tensor $\Psi$ \cite{paul} where each entry gives the number of
agents holding a particular combination of $S$ strategies.  We
assume $\Psi$ to be constant over the timescale for which
time-averages are taken. A single $\Psi$ `macrostate' corresponds
to many possible `microstates' describing the specific partitions
of strategies among the $N$ agents \cite{paul}. To allow for large
strategy spaces and large sets of global information, we consider
$\{R\}$ and $\mu(t)$ to be numbers on the line from
$R=1\rightarrow R_{\rm max}$, and from $\mu=1\rightarrow\mu_{\rm
max}$ respectively. For small strategy spaces, the subsequent
integrals can be converted to sums. Denoting the number of agents
choosing $-1$ ($+1$) as $N_{-1}(t)$ ($N_{+1}(t)$), the excess
number choosing $+1$ over $-1$ represents the inefficiency of the
system and is given by $D(t)=N_{+1}(t)-N_{-1}(t)$. In the context
of financial markets, $D(t)$ would be proportional to the
price-change representing the excess of demand over supply.
Similar analysis can be carried out for any function of $D(t)$,
$N_{+1}(t)$ and/or $N_{-1}(t)$, and time-cumulative value of these
quantities. Here we focus on $D(t)$ which is given exactly by:
\begin{equation} D(t)
\equiv \int_{R=1}^{R_{\rm max}} dR\  a_{R}^{\mu (t)}
n_{R}^{{\underline{S}(t);\Psi}}, \label{defD}
\end{equation}
where $\underline{S}(t)$ is the current score-vector denoting the
past performance of each strategy \cite{paul}. The combination of
$\underline{S}(t)$, $\Psi$ and the game rule (e.g. use strategy
with best or second-best performance to date) will define the
number of agents $n_{R}^{{\underline{S}(t);\Psi}}$ using strategy
$R$ at time $t$. The action $a_{R}^{\mu (t)}=\pm 1$ is determined
uniquely by ${\mu(t)}$.

In conventional many-body Physics, we are either interested in the
dynamical properties of $D(t)$, such as the equation-of-motion, or
its statistical properties \cite{mahan}. Here we focus on these
statistical properties: (i) the moments of the probability
distribution function (PDF) of $D(t)$ (e.g. mean, variance,
kurtosis) and (ii) the correlation functions that are products of
$D(t)$ at various different times $t_1=t$, $t_2=t+\tau$,
$t_3=t+\tau'$ etc. (e.g. autocorrelation). Numerical multi-agent
simulations typically average over time $t$ and then over
configurations $\{\Psi\}$. A general expression to generate all
such functions, is therefore {\small\begin{eqnarray}
&&D_P^{(\tau,\tau',\tau'',\dots)} \equiv \left\langle\left\langle
D(t_1) D(t_2) \dots D(t_P)\right\rangle_t \right\rangle_\Psi
\label{gen}\\ & = & \left\langle\left\langle
\int\dots\int_{\{R_i\}}^{R_{\rm max}}dR_1\ dR_2 \dots dR_P \
a_{R_1}^{\mu(t_1)} a_{R_2}^{\mu(t_2)} \dots a_{R_P}^{\mu(t_P)}
n_{R_1}^{\underline{S}(t_1);\Psi}
n_{R_2}^{\underline{S}(t_2);\Psi} \dots
n_{R_P}^{\underline{S}(t_P);\Psi} \right\rangle _{t} \right\rangle
_{\Psi} \nonumber \\ & \equiv & \int\dots\int_{\{R_i\}}^{R_{\rm
max}}dR_1\ dR_2 \dots dR_P \left\langle\left\langle
V^{(P)}(R_1,R_2,\dots,R_P;t_1,t_2,\dots,t_P)
n_{R_1}^{\underline{S}(t_1);\Psi}
n_{R_2}^{\underline{S}(t_2);\Psi}  \dots
n_{R_P}^{\underline{S}(t_P);\Psi} \right\rangle _{t} \right\rangle
_{\Psi} \nonumber
\end{eqnarray}}

\noindent where
\begin{equation} V^{(P)}(R_1,R_2\dots,R_P;t_1,t_2\dots,t_P)\equiv
a_{R_1}^{\mu(t_1)} a_{R_2}^{\mu(t_2)}  \dots a_{R_P}^{\mu(t_P)} \label{defV}
\end{equation} resembles a time-dependent, non-translationally invariant, {\em
$p$-body} interaction potential in ${\bf R}\equiv
(R_1,R_2\dots,R_P)$-space, between $p$ charge-densities
$\{n_{R_i}^{\underline{S}(t_i);\Psi}\}$ of like-minded agents.
Note that each charge-density $n_{R_i}^{\underline{S}(t_i);\Psi}$
now possesses internal degrees of freedom determined by
$\underline{S}(t)$ and $\Psi$. Since
$\{n_{R_i}^{\underline{S}(t_i);\Psi}\}$ are determined by the
game's rules, Eq. (\ref{gen}) can be applied to {\em any}
multi-agent game, not just MG. We focus here on moments of the PDF
of $D(t)$ where $\{t_{i}\} \equiv t$ and hence $\{\tau\}=0$.
Discussion of temporal correlation functions such as the
autocorrelation $D_{2}^{(\tau)}$ will be reported elsewhere. We
consider explicitly the variance $D_{2}$ to demonstrate the
approach, noting that higher-order moments such as $D_{4}$ (i.e.
kurtosis) which classify the non-Gaussianity of the PDF, can be
treated in a similar way. The potential $V^{(P)}$ is insensitive
to the configuration-average over $\{\Psi\}$, hence the mean is
given by \cite{note}:
\begin{equation} D_{1} = \int_{R=1}^{R_{\rm max}}dR \  \
\left\langle V^{(1)}(R;t)\
\left\langle n_{R}^{\underline{S}(t);\Psi} \right\rangle _{\Psi}\right\rangle
_{t}\ .
\end{equation} If the game's output is unbiased, the averages yield $D_1=0$.
This condition is not necessary -- one can simply subtract $D_1^2$ from the
right hand side of the expression for $D_2$ below -- however we will take
$D_{1}=0$ for clarity. The variance $D_2$ measures the fluctuations of $D(t)$
about its average value:
\begin{equation} D_2 = \int\int_{R,R^{\prime }=1}^{R_{\rm max}}dR dR^{\prime
}\  \left\langle V^{(2)}(R,R^{\prime};t)\
\left\langle n_{R}^{\underline{S}(t);\Psi} n_{R^{\prime
}}^{\underline{S}(t);\Psi}\right\rangle _{\Psi}\right\rangle _{t} \label{var}
\end{equation} where $V^{(2)}(R,R^{\prime};t)\equiv  a_{R}^{\mu (t)}
a_{R^{\prime }}^{\mu (t)}$ acts like a time-dependent, non-translationally
invariant, two-body interaction potential in
$(R,R')$-space.
Figure 2 illustrates a diagrammatic representation of $D_{P=1,2}$
in analogy with conventional many-body theory.

The effective charge-densities and potential will fluctuate in time.
It is reasonable to assume that the charge densities fluctuate around some mean
value, hence
$n_{R}^{\underline{S}(t);\Psi} n_{R'}^{\underline{S}(t);\Psi}=n_{R}
n_{R'}+\varepsilon_{R R'}^{\underline{S}(t);\Psi}(t)$ with mean $n_R n_{R'}$
plus a fluctuating term $\varepsilon_{R R'}^{\underline{S}(t);\Psi}(t)$. This
is a good approximation if we take $R$ to be a popularity-ranking (i.e. the
$R$th most popular strategy) or a strategy-performance ranking (i.e. the $R$th
best-performing strategy) since in these cases $n_{R}^{\underline{S}(t);\Psi}$
will be reasonably constant. For example, taking $R$ as a popularity-ranking
implies $n_{R=1}^{\underline{S}(t);\Psi}\geq n_{R=2}^{\underline{S}(t);\Psi}\geq
n_{R=3}^{\underline{S}(t);\Psi}\geq \dots$, thereby constraining the magnitude
of the fluctuations in the charge-density
$n_{R}^{\underline{S}(t);\Psi}$.  Hence
\begin{equation} D_2 = \int\int_{R,R^{\prime }=1}^{R_{\rm max}}dR dR^{\prime
}\  \left\langle V^{(2)}(R,R^{\prime};t)\ \left\langle n_{R}
n_{R'}+\varepsilon_{R R'}^{\underline{S}(t);\Psi}(t)
\right\rangle _{\Psi} \right\rangle _{t}\ .
\end{equation} We will assume that
$\varepsilon_{RR'}^{\underline{S}(t);\Psi}(t)$ averages out to
zero. In the presence of network connections between agents, there
can be strong correlations between these noise terms
$\varepsilon_{R R'}^{\underline{S}(t);\Psi}(t)$ and the
time-dependence of $V^{(2)}(R,R^{\prime};t)$, implying that the
averaging over $t$ should be carried out step-by-step as in Ref.
\cite{Lo}. For MG-like games without connections, the agents
cannot suddenly access larger numbers of strategies and hence
these correlations can be ignored. This gives
\begin{equation} D_2 =
\int\int_{R,R^{\prime }=1}^{R_{\rm max}}dR dR^{\prime }\  \left\langle
V^{(2)}(R,R^{\prime};t) \right\rangle _{t} \ n_{R} n_{R'}\ .\label{D2mid}
\end{equation} As in conventional many-body theory, the expectation value in Eq.
(\ref{D2mid}) can be `contracted' down by making use of the
equal-time correlations between $\{a^{\mu(t)}_{R}\}$. As is known
for MG-like games \cite{others,us,chau}, the complete strategy
space will contain strategies which have exactly the same
responses except for a few $\mu(t)$'s. This quasi-redundancy can
be removed by focusing on a reduced strategy space such that any
pair $R$ and $R'$  are either (i) correlated, i.e.
$a^{\mu(t)}_R=a^{\mu(t)}_{R'}$ for all (or nearly all) $\mu(t)$;
(ii) anti-correlated, i.e. $a^{\mu(t)}_R=-a^{\mu(t)}_{R'}$ for all
(or nearly all) $\mu(t)$;  (iii) uncorrelated, i.e.
$a^{\mu(t)}_R=a^{\mu(t)}_{R'}$ for half (or nearly half) of
$\{\mu(t)\}$ while $a^{\mu(t)}_R=-a^{\mu(t)}_{R'}$ for the other
half of $\{\mu(t)\}$. Hence one can choose two subsets of
$\Theta$, i.e. $\Theta=U\oplus \overline U$, such that the
strategies within $U$ are uncorrelated, the strategies within
$\overline U$ are uncorrelated, the anticorrelated strategy of
$R\in U$ appears in $\overline U$, and the anticorrelated strategy
of $R\in \overline U$ appears in $U$. We can therefore break up
the integrals in Eq. (\ref{D2mid}) into three parts: (i) $R'\sim
R$ (i.e. correlated) hence $\frac{1}{\mu_{\rm
max}}\int_{\mu=1}^{\mu_{\rm max}} d\mu a_{R}^{\mu} a_{R'}^{\mu}
=1$ and $\left\langle V^{(2)}(R,R^{\prime};t)\right\rangle
_{t}=1$. (ii) $R'\sim {\overline R}$ (i.e. anticorrelated) which
yields $\frac{1}{\mu_{\rm max}}\int_{\mu=1}^{\mu_{\rm max}} d\mu
a_{R}^{\mu} a_{ R'}^{\mu} =-1$. If all possible global information
values $\{\mu\}$ are visited reasonably equally over a long
time-period, this implies $\left\langle
V^{(2)}(R,R^{\prime};t)\right\rangle _{t}=-1$. For the MG, for
example, $\{\mu\}$ corresponds to the $m$-bit histories which
indeed are visited equally for small $m$. For large $m$, they are
not visited equally for a given $\Psi$, but are when averaged over
all $\Psi$. If, by contrast, we happened to be considering some
general non-MG game where the $\mu$'s occur with unequal
probabilities $\rho_\mu$, even after averaging over all $\Psi$,
one can simply redefine the strategy subsets $U$ and $\overline U$
to yield a generalized scalar product, i.e. $\frac{1}{\mu_{\rm
max}}\int_{\mu=1}^{\mu_{\rm max}} d\mu a_{R}^{\mu} a_{R'}^{\mu} \
\rho_\mu=-1$ (or $0$ in case (iii)). (iii) $R'\perp R$ (i.e.
uncorrelated) which yields $\frac{1}{\mu_{\rm
max}}\int_{\mu=1}^{\mu_{\rm max}} d\mu a_{R}^{\mu} a_{R'}^{\mu}
=0$ and hence $\left\langle V^{(2)}(R,R^{\prime};t)\right\rangle
_{t}=0$. Hence
\begin{eqnarray} D_2 & = & \int\int_{R,R^{\prime }=1}^{R_{\rm max}}dR
dR^{\prime }\  \left\langle V^{(2)}(R,R^{\prime};t)\right\rangle _{t}\ n_{R}
n_{R'}=
\int_{R=1}^{R_{\rm max}}dR \  (n_{R} n_{R}-n_{R} n_{\overline R})\nonumber \\ &
= &
\int_{R\in U} dR \  (n_{R} n_{R}-n_{R} n_{\overline R}+ n_{\overline R}
n_{\overline R}-n_{\overline R} n_{R})
\nonumber \\ &
 = &
\int_{R\in U} dR \  (n_{R} - n_{\overline R})^2 \ .
\label{final}
\end{eqnarray} Equation (\ref{final}) must be evaluated together with the
condition which guarantees that the total number of agents $N$ is conserved:
\begin{equation} N=\int_{R=1}^{R_{\rm max}}dR \  n_{R}\equiv \int_{R\in U} dR
\  (n_{R} + n_{\overline R}) \ .
\label{vol}
\end{equation} Equation (\ref{final}) has a simple interpretation. Since
$n_{R}$ and $n_{\overline R}$ have opposite sign, they act like two
charge-densities of opposite charge which tend to cancel: $n_{R}$ represents
a Crowd of like-minded people, while $n_{\overline R}$
corresponds to a like-minded Anticrowd who do exactly the {\em opposite} of the
Crowd. We have
effectively renormalized the charge-densities
$n_{R}^{\underline{S}(t);\Psi}$ and
$n_{R'}^{\underline{S}(t);\Psi}$ and their time- and position-dependent
two-body interaction $V^{(2)}(R,R';t)\equiv a_{R}^{\mu (t)} a_{R^{\prime
}}^{\mu (t)}$, to give two identical Crowd-Anticrowd `quasiparticles'
of charge-density $({{n_{R}}}-{{
{n_{\overline{R}}}}})$ which interact via a {\em time-independent} and {\em
position-independent} interaction term $V^{(2)}_{\rm eff}\equiv 1$.
This is
shown schematically in Fig. 2.
The different types of Crowd-Anticrowd quasiparticle in Eq.
(\ref{final}) do not interact with each other, i.e. $({{n_{R}}}-{{
{n_{\overline{R}}}}})$ does not interact with $({{n_{R'}}}-{{
{n_{\overline{R'}}}}})$ if $R\neq R'$. Interestingly, this situation could
{\em not} arise in a conventional physical system containing just two types of
charge (i.e. positive and negative).

A given numerical simulation will employ a given
strategy-allocation matrix (i.e. a given rank-$S$ tensor) $\Psi$.
As $R_{\rm max}$ increases from $1\rightarrow\infty$, $\Psi$ tends
to become increasingly disordered (i.e. increasingly non-uniform)
\cite{book,paul} since the ratio of the standard deviation to the
mean number of agents holding a particular set of $S$ strategies
is equal to $[({R_{\rm max}^S}-1)/N]^\frac{1}{2}$. There are two
regimes: (i) A `high-density' regime where $R_{\rm max}\ll N$.
Here the charge-densities 
$\{n_R\}$ tend to be large, non-zero values which monotonically
decrease with increasing $R$.  Hence the set $\{n_R\}$ acts like a
smooth function $n(R)\equiv \{n_R\}$.  (ii) A `low-density' regime
where $R_{\rm max}\gg N$. Here $\Psi$ becomes sparse with each
element $\Psi_{\rm R,R',R'',\dots}$ reduced to 0 or 1. The
$\{n_R\}$ should therefore be written as 1's or
0's in order to retain the discrete nature of the agents, and yet
also satisfy Eq. (\ref{vol}) \cite{book}. Depending on the
particular type of game, moving between regimes may or may not
produce an observable feature. In the MG, for example, $D_1$ does
not show an observable feature as $R_{\rm max}$ increases --
however $D_2$ does \cite{savit}. We leave aside the discussion as
to whether this constitutes a true phase-transition
\cite{others,paul} and instead discuss the explicit analytic
expressions for $D_2$ which result from Eq. (\ref{final}).  It is
easy to show that the mean number of agents using the $X$th most
popular strategy (i.e. after averaging over $\Psi$) is
\cite{book}:
\begin{equation} {{n_{X}}} = N \bigg[
\left( 1-\frac{(X-1)}{R_{\rm max}}\right) ^{S}-\left( 1-\frac{X}{R_{\rm max}}
\right) ^{S}\bigg] . \label{nav}
\end{equation}
The increasing non-uniformity in $\Psi$ as $R_{\rm max}$
increases, means that the popularity-ranking of $\overline R$
becomes increasingly independent of the popularity-ranking of $R$.
Using Eq. (\ref{nav}) with $S=2$, and averaging over all possible
$\overline R$ positions in Eq. (\ref{final}) to reflect the
independence of the popularity-rankings for $\overline R$ and $R$,
we obtain:
\begin{equation} D_2 ={\rm
Max}\bigg[\ \ {\frac{N^2}{3 R_{\rm max}}} \bigg(1-{R^{-2}_{\rm
max}}\bigg),\ \ N\bigg(1-\frac{N}{R_{\rm
max}}\bigg)\bigg]\label{lower}\ . \label{Dfin}
\end{equation}
The `Max' operation ensures that as $R_{\rm max}$ increases and
hence $\{n_R\}\rightarrow 0,1$, Eq. (\ref{vol}) is still satisfied
\cite{book}. Equation (\ref{Dfin}) underestimates $D_2$ at small
$R_{\rm max}$ (see Fig.~1) since it assumes that the rankings of
$\overline R$ and $R$ are unrelated, thereby overestimating the
Crowd-Anticrowd cancellation. By contrast, an overestimate of
$D_2$ at small $R_{\rm max}$ can be obtained by considering the
opposite limit whereby $\Psi$ is sufficiently uniform that the
popularity and strategy-performance rankings are identical. Hence
the strategy with popularity-ranking $X$ in Eq. (\ref{nav}) is
anticorrelated to the strategy with popularity-ranking $R_{\rm
max}+1-X$. This leads to a slightly modified first expression in
Eq. (\ref{Dfin}): $\frac{2 N^2}{3 R_{\rm max}} (1-R_{\rm
max}^{-2})$. Figure 1 shows that the resulting analytical
expressions reproduce the quantitative trends in the standard
deviation $D_{2}^{1/2}$ observed numerically for all $N$ and
$R_{\rm max}$, {\em and} they describe the wide spread in the numerical data
observed at small $R_{\rm max}$.

In summary, we have uncovered an explicit connection between
multi-agent games and conventional many-body theory. This should
not only help to bring multi-agent games closer to the Physics
community, but it should also help the Physics community step into
non-traditional areas of research where multi-agent simulations
are now being actively used.

\newpage

\begin{figure}
\includegraphics{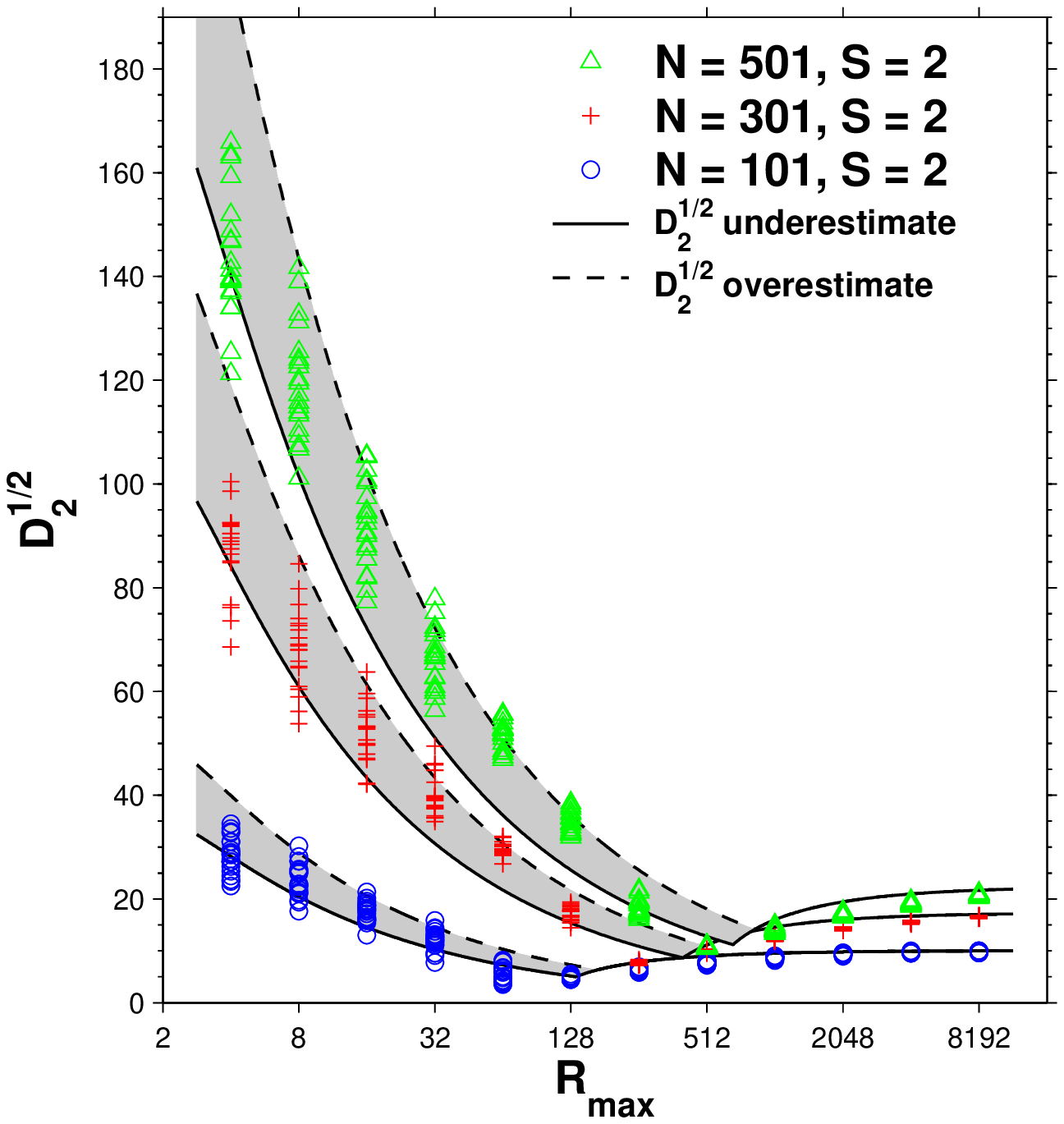}%
\caption{(Color online) Results for the standard deviation of fluctuations
$D_{2}^{1/2}$ in the Minority Game. Numerical results correspond to 20
different runs at each $N$ and $R_{\rm max}$. The theoretical curves
are generated using the analytic expressions in the text. The shaded area
bounded by the upper and lower curves shows our theoretical prediction of the
numerical spread for a given $N$.  In line with the original numerical results of
Ref.
\cite{savit}, we have chosen successive $R_{\rm max}$ tick-values to
increase by a factor of 4. }
\label{figure2}
\end{figure}

\begin{figure}
\includegraphics{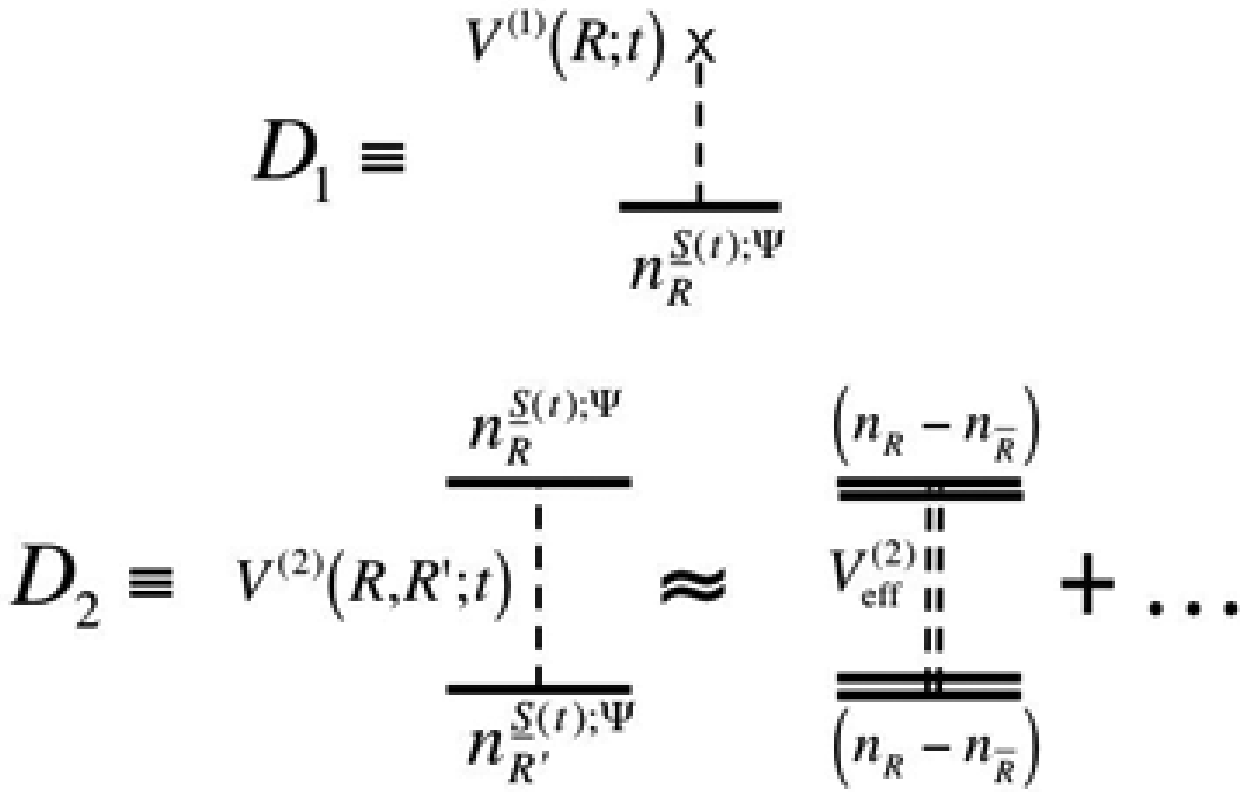}%
\caption{Diagrammatic equivalent (schematic) of some simple $p$th order
moments $D_P$ of the multi-agent output variable
$D(t)$. Top: $p=1$ is the mean $D_1$. Bottom: $p=2$ is the variance
$D_2$. Also shown is an approximate expansion for $D_2$ as represented by Eq.
(\ref{final}). Higher-order terms, which are neglected in Eq.
(\ref{final}), would correspond to residual
correlations between distinct Crowd-Anticrowd quasiparticles.}
\label{figure2}
\end{figure}


\begin{thebibliography}{99}
\bibitem{note} We interchange the order of the $\Psi$ and $t$-averaging
over the product of the $n_R$'s. Numerical simulations
suggest this is valid for the systems of interest.
\end{thebibliography}
\end{document}